***Nanocryotron-driven Charge Configuration Memristor***

Anže Mraz*[1,2,3], V.V. Kabanov[1], Rok Venturini[4], Damjan Svetin[1], Viktoriia Yursa[2], Igor Vaskivskyi[1,2], Bor Brezec[3], Tevž Lotrič[5], Matic Merljak[3], Jan Ravnik[4], Dimitrios Kazazis[4], Simon Gerber[4], Yasin Ekinci[4], Mihai Gabureac[6#] and D. Mihailovic[1,2,5]

*[1] Jozef Stefan Institute, Jamova 39, SI-1000 Ljubljana, Slovenia*
*[2]CENN Nanocenter, Jamova 39, SI-1000 Ljubljana, Slovenia*
*[3]Faculty for Electrical Engineering, University of Ljubljana, Tržaška 25, SI-1000 Ljubljana, Slovenia*
*[4]PSI Center for Photon Science, Paul Scherrer Institute, CH-5232 Villigen PSI, Switzerland*
*[5]Faculty for Mathematics and Physics, University of Ljubljana, Jadranska 19, SI-1000 Ljubljana, Slovenia*
*[6]Department of Physics, ETH Zurich, CH-8093 Zurich, Switzerland*
*Email: anze.mraz@ijs.si

**Cryo-computing – both classical and quantum, is severely limited by the absence of a suitable cryo-memory[1–3]. The challenge both in terms of energy efficiency and speed have been known for decades, but so far conventional technologies have not been able to deliver adequate performance. Here we present a novel *non-volatile* memory device which incorporates a superconducting nanowire[4] and an all-electronic charge configuration memristor (CCM)[5,6] based on switching between charge-ordered states in a layered dichalcogenide material. We investigate the time-dynamics and current-voltage characteristics of such a device fabricated using a NbTiN nanowire and a $1T$-$TaS_2$ CCM. The observed dynamical response of the device is faithfully reproduced by modelling of the superconducting order parameter showing versatility of application[7]. The inherent ultrahigh energy efficiency and speed of the device, which is compatible with single flux quantum logic, leads to a promising new memory concept for use in cryo-computing and quantum computing peripheral devices.**

## Introduction

The future of conventional computing is currently severely limited by processor heat dissipation[8] and relatively slow advances in memory performance. Superconductor-based computing has been heralded as an obvious solution to heat dissipation[2,8], and energy-efficient single-flux-quantum (EESFQ) circuits have recently been developed that also offer very high operating speed and efficiency[9–11]. Concurrently, superconductor-based quantum computing in various guises has demonstrated impressive advances in recent years. A particularly important advance is the possibility of driving qubits with programmable trains of single flux quantum pulses[12]. However, the development of this, and other cryo-computing (CC) technology is inhibited by the lack of a suitable fast and energy-efficient cryomemory[2,3]. The development of such a memory is not only important for SFQ computers, but also for superconducting (SC) quantum computing, where it allows for scalable control and error-correction circuits to be placed near the quantum processor[13], reducing the need for numerous noise- and heat- generating control cables that lead to decoherence and introduce cosmic-ray and other ambient noise[14]. A number of different approaches to introduce memory into a CC environment have recently been investigated with various degrees of success, including cryo-CMOS[15,16], SC memory[1] and magnetic memory[17,18]. Most recent purely-SC memories have relied on low kinetic inductance niobium wiring[19] to reduce the overall element size, but at present the problem of fast, scalable, energy efficient memory devices is still not solved.

Here we present a new approach to the cryo-memory problem based on $1T$-$TaS_2$ charge configuration memristors (CCM) devices that have recently been shown to operate down to milli-Kelvin temperatures[6]. The devices rely on ultrafast non-volatile[20] switching between electronic domain configurations within the charge-ordered material in response to external electromagnetic stimuli[5,21]. The energy required for switching has been shown to be < 2.2 fJ/bit, while the duration of electrical pulses for control can range from ms to less than 2 ps[22]. This makes CCMs good candidates for ultrafast low-power non-volatile memory applications demanded by EESFQ circuits. However, the desired switching energies for use in SFQ circuits is set by the small magnitude of the SFQ pulse $\phi_0 = 2$ mVps, which carries the bits of data. For typical currents, the energy of such pulses is $E_{\phi_0} \sim 10^{-20}$ J. Considering that CCM devices typically require $\sim 500$ mV for switching, amplification is needed to drive them by SFQ devices. The amplifiers should be fast, require a small amount of energy, and allow control of CCM by single SFQ pulses. A SC nanowire (NW) cryotron (nTron) device, based on gate-controllable switching between the normal and SC state of the NW, and which can be driven by SFQ pulses, was recently reported[4]. The nTron was shown to be capable of driving conventional electronics, connecting SFQ circuits to CMOS memories[23,24]. Unfortunately, because of the high dissipation, CMOS memories are too inefficient for practical use in cryo-computing circuits, and more efficient devices such as CCMs are desirable. The nTron can have an output impedance in the ON state of $R_N > 10$ kΩ, and produce $> 1$ V at the output[4], which is more than sufficient for driving CCMs.

Here we investigate SFQ-switching time-dynamics and current-voltage characteristics of CCM-shunted SC NWs and, by extension, CCM-shunted nTrons shown in Fig. 1. To gain an understanding of the operation of such a hybrid 'parallelotron' (or 'pTron') device, we model the switching behaviour using a Josephson junction model previously applied to model the phase slip dynamics in capacitively and resistively shunted SC NWs[7]. We show that memory operation of the hybrid pTron device can be achieved using conventional enabling bias currents and typical SFQ trigger pulses. We investigate the dynamical behaviour on the ps timescale and discuss the speed limitations of pTron devices due to SC NW quasiparticle and thermal diffusion, as well as other circuit constraints. The hybrid pTron device is realized by fabricating a NbTiN nTron with a single-cell CCM based on a $1T$-$TaS_2$ exfoliated single crystal. Measurements of the device's performance in comparison with model calculations show remarkable agreement, highlighting the desirable parameter range for optimum operation of pTron devices.

## The dynamical response of a pTron.

The properties of a resistively and capacitively shunted Josephson junction (RCSJ) model for describing the properties of SC NWs were investigated in detail by Brenner *et al.*[7], who showed two different regimes of operation as the NW approaches the critical current $J_c$, depending on the magnitude of the shunt resistance $R_s$. In one regime, dissipation is caused by the appearance of fluctuating phase slip centres (PSCs) in which the phase $\phi$ of the SC order parameter $\psi = \psi_0 e^{-\phi}$ locally exhibits a shift of $2\pi$ which corresponds to the movement of a quantum of magnetic flux $\Phi_0 = h/2e$ from one side of the wire to the other. In the other, quasiparticles (QP) that are a result of pair-breaking (PB) dissipate energy in a process similar to Joule heating. The shunt resistance $R_s$ represented by CCM devices is typically $10\sim 20$ kΩ in the off state ($R_{\mathrm{HI}}^{\mathrm{CCM}}$), and $200\sim 400$ Ω in the on state ($R_{\mathrm{LO}}^{\mathrm{CCM}}$), so the contrast in resistance is high enough that in principle the device can be used to switch the system between different voltage-current ($V - J$) curves, or even between the PB and PSC regimes using single SFQ pulses. In the context of the pTron, the shunt CCM device is considered as a current-controlled switchable resistor, whose switching currents are typically 50~200 uA, corresponding to voltages across the devices of $V_{\mathrm{CCM}} = 0.5\sim 1$ V, depending on the device. A more detailed description of the CCM's $V-J$ characteristics is presented in the supplemental information (SI-Fig. 1).

In narrow SC NW channels whose widths $w$ are comparable or smaller than the SC coherence length $\xi_s$, the critical current $J_c$ is determined by PB, and the appearance of rapidly fluctuating PSCs along the NW. For example, 100 nm wide NbTiN NWs show characteristic PSC steps in the $V-J$ curves, particularly at intermediate and high temperatures as $T \rightarrow T_c$[25]. Similar behaviour was observed in $\delta$-MoN nanowires[26,27]. The crossover between the PSC fluctuation and the PB regimes is described quite well by the McCumber Josephson junction model[7,28]. Here we extend this model to the pTron by including a CCM shunt. We consider the behaviour of a SC NW as well as an nTron, which is effectively a SC NW driven by an external gate current through a choke constriction whose function is to trigger a local hot-spot in a NW, current biased just below $J_c$. The circuit is shown in Fig. 1c, where $R_{\mathrm{CCM}}$ is the CCM resistance, $C$ is the circuit capacitance, $R_{\mathrm{N}}$ is the normal state resistance of the nanowire and $J_{\mathrm{b}}$ is the bias current. $j_{\mathrm{choke}}$ represents the gate current that is applied to the choke constriction that initiates switching. Following Ref.[7], the model describes the time-dynamics of the phase $\phi$ of the SC order parameter $\psi = \psi_0 e^{i\phi}$ within the SC NW. By applying Kirchhoff's laws and the Josephson's relations $J = J_{\mathrm{C}} \sin(\phi)$ and $V = \frac{\hbar}{2e}\frac{d\phi}{dt}$ , the equation of motion becomes:

$$Q^2 \frac{d^2\phi}{dt'^2} + \frac{d\phi}{dt'} + \sin\phi = J_{\mathrm{b}}, \qquad (1)$$

where $J_{\mathrm{b}}$ is the current through the SC NW channel, $J_c$ is the critical current of the NW, $Q = R_{\mathrm{T}}\sqrt{2eJ_cC/\hbar}$ is the quality factor and $R_{\mathrm{T}} = \frac{R_{\mathrm{N}}R_{\mathrm{CCM}}}{R_{\mathrm{N}}+R_{\mathrm{CCM}}}$ is the parallel resistance of $R_{\mathrm{N}}$ and $R_{\mathrm{CCM}}$, which are the normal resistance of the NW and the shunt resistance corresponding to the CCM device, respectively. Time $t'$ is measured in units of $t' = \left(\frac{2eJ_cR_{\mathrm{T}}}{\hbar}\right)t$.

Alternatively, the behaviour of the SC NW can also be described in terms of time-dependent Ginzburg-Landau equation (TDGL)[29–31]. Eq. (1) neglects all fluctuations of the superfluid density and considers only dynamics of the phase $\phi$. Therefore, it represents the limit of TDGL in the limit of large $u \gg 1$, where $u$ is the ratio of the relaxation time of the superfluid density to the phase relaxation time. The behaviour of the nanowire within TDGL approach was discussed in Ref.[30,31]. Similar as for the TDGL model, for $J_{\mathrm{b}} < J_c$, the RCSJ model in Eq. (1) has two steady state solutions, defined by the condition $\sin\phi = J_{\mathrm{b}}/J_c$. One of these solutions is stable until $J_{\mathrm{b}} = J_c$ and the second is always unstable. When the current reaches the critical current value these two steady state solutions collide and a periodic-in-time stable limit cycle solution emerges. This solution is characterized by the finite voltage in the circuit and Ohmic losses. Without noise the steady state solution becomes unstable exactly at the depairing critical current. The presence of the current noise[7] leads to a stochastic transition to the normal state at a smaller current which is usually called the switching current $J_{\mathrm{sw}}$. The periodic limit cycle solution is stable also below critical current. At the retrapping current $J_{\mathrm{ret}} < J_c$ the limit cycle collides with the unstable steady state solution and loses it stability[30,31]. As it was demonstrated in Ref.[30,31], the stability of different solutions of both the TDGL model and the RCSJ model described in Eq. (1) depends on the shunt resistance. Decreasing the shunt resistance leads to decreasing the difference between switching and retrapping currents.

We proceed by first considering the behaviour of a SC NW without a CCM. Figures 2a and b show the temporal evolution of Eq. (1) for $\phi(t')$ and the resulting voltage $V = J_cR_{\mathrm{T}}\frac{\mathrm{d}\phi}{\mathrm{d}t'}$ in the NW after application of current $J_{\mathrm{b}}$ at time $t' = 0$. As expected, below the NW switching current ( $J_{\mathrm{b}} < J_{\mathrm{sw}}$), the application of the current causes small damped oscillations of $\phi(t')$ and $V$ with a frequency $\nu = 2eJ_{\mathrm{b}}R_{\mathrm{T}}/\hbar$, which revert to zero with a time constant $\tau_{R_{\mathrm{T}}C} = R_{\mathrm{T}}C$ (Fig. 2a). When $J_{\mathrm{b}} > J_{\mathrm{sw}}$, we observe a transition to the normal state as the phase $\phi(t')$ increases with time and a finite $V$ appears across the NW which settles to a constant value within a time defined by $\sim\tau_{R_{\mathrm{T}}C}$ (Fig. 2b).

Introducing the CCM shunt resistance into Eq. (1), similar behaviour as in Fig. 2b is observed for voltage $V$ until the CCM switching threshold is reached, given by $J_{\mathrm{W}}^{\mathrm{CCM}} = V/R_{\mathrm{T}}$. Thereupon, the device switches to a low resistance state with a time given by $\tau_{\mathrm{LO}} = R_{\mathrm{T}}C$ ($\simeq$ 0.4 ps) (Fig. 2c) with the CCM in the $R_{\mathrm{LO}}^{\mathrm{CCM}}$ state. The pTron then remains either in a low-resistance state or the SC state, depending on the value of $R_{\mathrm{LO}}^{\mathrm{CCM}}$ as shown in Fig. 2c, where the parameter values are listed in the caption.

### Switching with $\phi_0$ pulses.

The ultimate goal is to perform memory operation on a pTron with single SFQ-like pulse. For this purpose, an ultrashort ($\sim$1 ps) SFQ write (W) pulse $j_{\mathrm{choke}}(t)$ is applied across a current biased NW as indicated in Fig. 1. In practical nTron devices the pulse is applied in the middle of the NW through a third terminal called the choke[4]. The pulse causes the SC of the choke to break and form a hotspot, which spreads into the channel, effectively decreasing the critical current of the channel[4]. To model the role of the choke, a pulsed bias current $J_{\mathrm{b}}(t)$ is applied synchronously across the NW, such that the total current through the NW is $J = J_{\mathrm{b}} + j_{\mathrm{choke}}$ (Fig. 1). The resulting switching is shown in Fig. 2d for $j_{\mathrm{choke}} = 5$ μA, $J_{\mathrm{b}} = 87$ μA and $\tau_{\mathrm{choke}} = 1$ ps. The bias current is in the form of a 100 ps pulse seen in the top panel. Other parameter values given in the figure caption are from typical nTron devices in the literature[4]. The choke current $j_{\mathrm{choke}} = 5$ μA corresponds to one flux quantum $\phi_0$ of width $\tau_{\mathrm{w}} = 1$ ps and amplitude of 2 mV converted to a current by a $R_{\mathrm{ser}} = 400\ \Omega$ resistor in series. The values of $j_{\mathrm{choke}}$ and $j_{\mathrm{b}}$ are very close to the reported nTron choke switching and channel currents of $\sim$7 μA and 30$\sim$120 μA respectively[4,23].

The switching dynamics for different magnitude $j_{\mathrm{choke}}$ is shown in Fig. 2e. With appropriate bias, the pTron switching threshold is < 5 μA, which means that it can easily be switched with a single SFQ pulse. Note that the resulting LO state of the pTron can be either SC or in a low-resistance state, depending on the value of $R_{\mathrm{LO}}^{\mathrm{CCM}}$ as shown in Fig. 2c.

### $V-J$ curves for writing, erasing and reading data

Having considered the switching dynamics for short SFQ-like pulses, we examine the $V-J$ characteristics of the NW and the pTron for the W, E and read (R) cycles calculated using Eq. (1). In Fig. 3a we show the NW $V-J$ direct-current (DC) curves for different shunt resistances $R_s$, illustrating the expected cross-over from the hysteretic regime with high $R_s$ to the non-hysteretic regime for low $R_s$. The switching and retrapping currents are also indicated in Fig. 3a. In the calculation we use $C = 1$ fF, and $J_{\mathrm{C}} = 120$ μA. With higher values of $R_s\sim 400\ \Omega$, and consequently high $Q\sim 300$, the Josephson term $\sin\phi$ in Eq. (1) is irrelevant. A smaller value of $R_s$ with a smaller $Q\sim 7$ as in Brenner *et al.*[7] leads to the PSC fluctuation regime as the oscillatory $\sin\phi$ becomes relevant. For our purposes, to reach sufficiently high voltages required by the CCM, it is convenient to operate the NW in the hysteretic regime.

For the $V-J$ curve calculation of Fig. 3a we have used the same initial conditions as in Ref.[7]: $\phi(0) = d\phi(0)/dt = 0$ for increasing current and $\phi(0) = d\phi(0)/dt = 1$ for decreasing current. The switching current in that case corresponds to the value of $J$ for which the point $\phi(0) = d\phi(0)/dt = 0$ of phase space moves from the basin of attraction of the stable steady state solution $\phi = \arcsin(J/J_c)$, to the basin of attraction of the stable limit cycle. Another choice of initial condition when the initial values of phase and its derivative for the new value of current are set to be equal to the final values of the phase and its derivative for the previous value of current[30,31] leads to a different $V-J$ curve, where the switching current is equal to the critical current $J_c$. This happens because in this case the initial conditions are always in the basin of attraction of the stable steady-state solution. Considering that initial conditions are defined by the procedure of experimental measurements, the $V-J$ curve of the device will be dependent on how the measurement is done. In a realistic situation the operation of a memory device is further complicated due to the switching current being dependent on external conditions, such as pulse shape and current noise. All this implies that there is some uncertainty in definition of the switching and retrapping currents for the SC NW[30,31]. However, it also makes it possible to tune the switching current by tailoring the pulse shape.

**The Write (W) process** for a pTron is shown in Fig. 3b. In the initial state the NW, and by extension the pTron, is superconducting. Increasing $J_{\mathrm{b}}$ above the switching current for a pTron in HI state ($J_{\mathrm{sw}}^{\mathrm{HI}}$) causes the pTron to switch to the normal state and a voltage appears across it with a value determined by the parallel resistance of the normal state resistance $R_{\mathrm{N}}$ of the nTron and the CCM shunt resistance $R_{\mathrm{HI}}^{\mathrm{CCM}}$. At this point $J_{\mathrm{b}}$ is divided between the NW and CCM branches in the ratio of their resistances. As $J_{\mathrm{b}}$ increases and the W threshold $J_{\mathrm{W}}^{\mathrm{CCM}}$ in the CCM's branch is reached, the CCM switches from $R_{\mathrm{HI}}^{\mathrm{CCM}}$ to $R_{\mathrm{LO}}^{\mathrm{CCM}}$, resulting in the overall drop in resistance of the pTron ($R_{\mathrm{HI}}^{\mathrm{pTron}} \rightarrow R_{\mathrm{LO}}^{\mathrm{pTron}}$) and in a voltage drop observed in the $V-J$. As $J_{\mathrm{b}}$ is decreased, we follow the dashed line which eventually retraps to the SC state. The device parameters used for the calculation shown in Fig. 3b are typical values from the literature for the NW (or nTron) and CCM[4,6,23] and are given in the figure caption. The plots in Fig. 3b are for three different values of $R_{\mathrm{N}} = 12\ \mathrm{k\Omega}$ (blue), $16\ \mathrm{k\Omega}$ (orange) and $20\ \mathrm{k\Omega}$ (green).
Even though $V-J$ curves in Fig. 3b are simulated using a DC current sweep, we can estimate the upper limit of the energy dissipated in the case of a pulsed W operation, as seen in Fig. 2d. For the green curve in Fig. 3b, $J_{\mathrm{W}}^{\mathrm{pTron}} \sim 90\ \mu\mathrm{A}$, $R_{\mathrm{N}}$=20 kΩ, $R_{\mathrm{HI}}^{\mathrm{CCM}}$= 8.5 kΩ and we assume the length of the pulse $\tau$ =100 ps. Writing energy to switch from HI to LO state is then $E_{\mathrm{W}} = \left(J_{\mathrm{W}}^{\mathrm{pTron}}\right)^2 R_{\mathrm{T}}\tau_{\mathrm{W}} \approx 5$ fJ, where $R_{\mathrm{T}} = R_{\mathrm{N}}R_{\mathrm{HI}}^{\mathrm{CCM}}/(R_{\mathrm{N}} + R_{\mathrm{HI}}^{\mathrm{CCM}})$ is the parallel resistance. Besides the energy supplied by the bias current, we also consider the small SFQ pulse through the choke that triggers the NW. Using resistance of the choke $R_{\mathrm{choke}} = 400\ \Omega$, $\tau_{\phi_0}$=2 ps and $J_{\mathrm{choke}}$=5 μA, gives $E_{\mathrm{choke}} \approx 2 \times 10^{-21}$ J. The dissipation of the choke is thus very small compared to the rest of the device.

**The Read (R) process.** To read the state of the pTron, the required current range is $J_{\mathrm{SW}}^{\mathrm{HI}} < J_{\mathrm{R}} < J_{\mathrm{W}}^{\mathrm{pTron}}$ as shown in Fig. 3c. Note that in the present case a narrow region between $60\ \mu\mathrm{A}$ and $68\ \mu\mathrm{A}$ exists, where the devices is superconducting in the $R_{\mathrm{LO}}^{\mathrm{pTron}}$ state, while having a finite voltage in the $R_{\mathrm{HI}}^{\mathrm{pTron}}$ state. Therefore, the optimal read current $J_{\mathrm{R}}^{\mathrm{opt}}$ would be somewhere in that region, where only reading of the HI state would produce dissipation, while reading of the LO state would be completely lossless. The width of this region corresponds to the difference of switching currents in the LO and HI states, which can be tuned by the nTron and CCM parameters. The maximum read energy is $E_{\mathrm{R}} = J_{\mathrm{R}}^2 R_{\mathrm{T}}\tau_{\mathrm{R}}$, which for the device with $R_{\mathrm{N}} = 20\ \mathrm{k\Omega}$ is in the range $2\ \mathrm{fJ} < E_{\mathrm{R}} < 3\ \mathrm{fJ}$.

**The Erase (E) process**. To revert from $R_{\mathrm{LO}}^{\mathrm{pTron}}$ to $R_{\mathrm{HI}}^{\mathrm{pTron}}$, the bias current of the channel is increased to approximately $2.2\ J_{\mathrm{c}}$ to reach the CCM erase threshold as shown in Fig. 3d. Note that the resistive part of the E cycle is quite similar to that of the CCM on its own as shown in SI-Fig. 1b. The E process of the pTron is governed predominantly by the erase threshold of the CCM. The largest amount of energy is required for the E operation, which is determined by thermal characteristics of the device. In practice, erase operations are performed in bulk, where speed and dissipation are not of primary importance.

**pTron measurements**

The nTrons (example shown in Fig. 4a) were fabricated by etching a sputter-deposited NbTiN SC thin film on Si substrate (more details in Methods). To form the hybrid pTron device, CCM and nTron are connected in parallel using a low-temperature multiplexer circuit, which allows multiple devices to be measured *in-situ* (Fig. 4b). Shielding and filtering is extremely important to reduce the effect of external noise-induced switching of the nTron. Noise is especially prevalent when it's coupled to the choke, which is very sensitive to electrical fluctuations due its small cross-section. However, to test the fast signal response, we omit the use of low-pass filters, which means that in these experiments the SC NW switching was triggered by current pulses across the channel, rather than SFQ-level picosecond pulses through the choke. SFQ-triggered operation of nTrons has already been previously demonstrated, and is not included in the present work[23].
The $V-J$ characteristics for write (W) and erase (E) processes of the pTron device using current pulses ($J_{\mathrm{b}}$) are shown in Fig. 4c and d, respectively. Note that the experiment is performed in a two-contact configuration, which is the reason for the small resistance at small current values (i.e., below the switching current) observed in Fig. 4. For the W process in Fig. 4c, initially the CCM (and by extension the pTron) is in the HI state. At low $J_{\mathrm{b}}$ pulse amplitude ($\tau_{\mathrm{b}}^{\mathrm{W}}$ =50 μs), the pTron is in the SC state and the voltage across it is zero. As $J_{\mathrm{b}}$ sweeps over the switching current $J_{\mathrm{sw}}^{\mathrm{HI}}$ (in this case $\sim 315\ \mu\mathrm{A}$), a voltage appears across the pTron, and we follow the green curve as $J_{\mathrm{b}}$ increases further. When $J_{\mathrm{b}}$ reaches the W threshold for the pTron, it switches from the HI state to LO state, observed by the drop in voltage. Measuring the resistance of the CCM element confirms that the device switched from $R_{\mathrm{HI}}^{\mathrm{CCM}} = 22\ \mathrm{k\Omega}$ to $R_{\mathrm{LO}}^{\mathrm{CCM}} = 460\ \Omega$. Upon decreasing $J_{\mathrm{b}}$ back to zero, we follow the orange curve and notice a transition back to the SC state at the retrapping current $J_{\mathrm{ret}}^{\mathrm{LO}} \sim 330\ \mu\mathrm{A}$.
For the E process shown in Fig. 4d the current sweeping procedure is the same, only that the pTron is initially in the LO state and longer pulses of $\tau_{\mathrm{b}}^{\mathrm{E}}$ =1 ms are used. At low $J_{\mathrm{b}}$ values, the voltage across the pTron is zero, right until we reach $J_{\mathrm{sw}}^{\mathrm{LO}} \sim 330\ \mu\mathrm{A}$ when a voltage appears. As $J_{\mathrm{b}}$ increases we follow the orange curve until we reach the E threshold for the pTron, at which point the pTron erases from the LO back to the HI state and we notice a jump in voltage. Measuring the resistance of the CCM element confirms that the device switched from $R_{\mathrm{LO}}^{\mathrm{CCM}} = 460\ \Omega$ to $R_{\mathrm{HI}}^{\mathrm{CCM}} = 26.7\ \mathrm{k\Omega}$. As $J_{\mathrm{b}}$ is decreased to zero we follow the green curve through a transition back to the SC state at $J_{\mathrm{ret}}^{\mathrm{HI}} \sim 315\ \mu\mathrm{A}$.
Figs. 4e-h show the read (R) operation in real time. By observing the voltage waveform across the pTron with an oscilloscope during the R operation, we can distinguish different resistance values of the pTron depending on the state of the pTron and the R current $J_{\mathrm{R}}$. In the presented case in Fig. 4e we have a pTron which is either in the HI or LO state, as seen by two different $V-J$ curves with different switching currents ($J_{\mathrm{sw}}^{\mathrm{LO}} > J_{\mathrm{sw}}^{\mathrm{HI}}$) due to different values of the CCM shunt resistance. Fig. 4f presents a case where a R pulse with an amplitude of $J_{\mathrm{R}} = 400\ \mu\mathrm{A}$ ($\tau_{\mathrm{R}} = 50\ \mu\mathrm{s}$) is too low to read the state of the pTron correctly as it is smaller than the switching currents for both the HI and LO state ($J_{\mathrm{R}} < J_{\mathrm{sw}}^{\mathrm{LO}}, J_{\mathrm{sw}}^{\mathrm{HI}}$). At a higher R current amplitude of $J_{\mathrm{R}} = 470\ \mu\mathrm{A}$ ($J_{\mathrm{sw}}^{\mathrm{HI}} < J_{\mathrm{R}} < J_{\mathrm{sw}}^{\mathrm{LO}}$) seen in Fig. 4g, the value for the LO state is the same as before (i.e., zero), but for the HI state we read a non-zero value which corresponds to the voltage in the HI state $V-J$ curve in Fig. 4e. If R current is increased further so that $J_{\mathrm{sw}}^{\mathrm{LO}}, J_{\mathrm{sw}}^{\mathrm{HI}} < J_{\mathrm{R}}$ (e.g., $J_{\mathrm{R}} = 530\ \mu\mathrm{A}$), we can read a distinct non-zero voltage value for both the HI and LO state, as seen in Fig. 4h. This means that a successful R operation can be performed with any $J_{\mathrm{R}}$ current that satisfies $J_{\mathrm{sw}}^{\mathrm{HI}} < J_{\mathrm{R}} < J_{\mathrm{W}}^{\mathrm{pTron}}, J_{\mathrm{E}}^{\mathrm{pTron}}$. The optimal R operation is performed in a specific current range marked with a green shaded area in Fig. 4e, where only reading of the HI state produces dissipation, while reading of the LO state is completely lossless, as it was also predicted in the numerical simulations (see Fig. 3c). In this range, the readout contrast $R_{\mathrm{HI}}^{\mathrm{pTron}}/R_{\mathrm{LO}}^{\mathrm{pTron}}$ is infinite.

**Discussion**

The desired operating regime of the SC NW is dictated by the W voltage of the CCM and its HI state resistance $R_{\mathrm{HI}}^{\mathrm{CCM}}$. Here, the typical values of $V_{\mathrm{CCM}} = 0.3 - 0.6$ V and $R_{\mathrm{HI}}^{\mathrm{CCM}} = 2\sim10$ k$\Omega$, keep the SC NW in the hysteretic regime. If the CCM switches to $R_{\mathrm{LO}}^{\mathrm{CCM}}$, then the SC NW can either remain in the PB regime or switches over to the PSC fluctuation regime, depending on the value of $R_{\mathrm{LO}}^{\mathrm{CCM}}$. In our current device, with most typical value of $R_{\mathrm{LO}}^{\mathrm{CCM}} \simeq 400\ \Omega$, the NW remains in the PB regime (Fig. 3b). The use of smaller CCM devices with smaller $R_{\mathrm{HI}}^{\mathrm{CCM}}$ and $V_{\mathrm{W}}^{\mathrm{CCM}}$ and/or smaller $R_{\mathrm{LO}}^{\mathrm{CCM}}$ would enable the device to switch between the PB and PSC regimes, which would have the advantage that in the LO state the pTron is superconducting, exhibiting an infinite $R_{\mathrm{HI}}/R_{\mathrm{LO}}$ ratio.

For unshunted NWs, or if $R_{\mathrm{s}}$ is very large, the retrapping and switching currents are substantially different. The NW behaviour is governed primarily by normal state Joule heating as a result of quasiparticle recombination and thermalization processes involving phonons that provide a mechanism for eventual thermalization of the NW. For low shunt resistances, the system resistance is primarily determined by PSC[7,26] and Joule heating becomes less important. The difference between switching and retrapping currents becomes small[7,30,31], and switching current reaches ultimate value of Bardeen's prediction for equilibrium depairing current[7].

Several factors play a role in the speed of the pTron which also have bearing on the applicability of the RCSJ model and also the eventual usefulness of the device. Within the model the characteristic response time is controlled by $R_{\mathrm{T}}$ and the circuit capacitance $C$, which in the present example gives an $RC$ constant $0.4\ \mathrm{ps} < \tau_{R_{\mathrm{T}}C} < 6\ \mathrm{ps}$. In comparison, the CCM intrinsic switching speed is as fast as $\sim 0.45$ ps[32], so this in principle should not limit overall device speed. In the pair-breaking regime, the QP dynamics in the SC NW is governed by QP recombination by optical phonon emission and the subsequent decay of high-energy optical phonons to low-energy dispersive acoustic phonons that carry heat out of the NW volume and eventually escape into the substrate. These processes, as well as phonon reabsorption, are described by Rothwarf-Taylor equations[33] in which the acoustic phonon escape out of the device is the rate-limiting step. These thermalization processes in NbN have been studied in detail within the context of Rothwarf-Taylor equations[34]. Optical pump-probe experiments directly measuring the transition from the SC to the normal state report this to occur within <10 ps in NbN sputtered films[34]. The phonon escape time $\tau_{\mathrm{esc}}$ was reported to occur on a timescale $\sim 100$ ps in a 15 nm thick NbN film[34], which is also discussed by MacCaughan et al. in the NW channel of an nTron device[4]. This is significantly longer than the phonon escape time in the ballistic limit for acoustic phonons generated by the quasiparticle recombination process can be made. (Assuming a velocity of sound $v_{\mathrm{s}} \simeq 5$ nm/ps, for a film thickness $\theta \simeq 15$ nm, a lower limit for the escape time is $\tau_{\mathrm{esc}} = \frac{\theta}{v_{\mathrm{s}}} \simeq 3$ ps.) Clearly the phonon escape process that determines the nTron dynamics is not ballistic, but diffusive[4,34]. In order to reach high operating speeds, the phonon escape timescale can be optimized either by varying film thickness and/or acoustic impedance matching at the interface, for example by including a buffer layer between the SC film and the substrate[35].

Finally, let us compare the QP recombination and diffusion timescales with the response time $\tau_{\mathrm{K}} = L_{\mathrm{K}}/R_{\mathrm{T}}$ determined from the kinetic inductance $L_{\mathrm{K}} = \frac{m_e}{2n_e e^2}\frac{l}{A}$. Here $l$ is the length of the wire, $A$ is the device cross-section and $n_e$ is the electron density. Using values for a typical device $l = 1\mu\mathrm{m}$, $A = w \times \theta\ \mathrm{nm}^2$, $n_e = 10^{24}\ \mathrm{m}^{-3}$, $L_{\mathrm{K}} \simeq 47$ nH, where $\theta = 15$ nm is the NbTiN film thickness and $w = 25$ nm is the width, which together with $R_{\mathrm{T}} \simeq 6$ k$\Omega$ gives an estimated $\tau_{\mathrm{K}} \simeq 8$ ps. Kinetic inductance thus does not appear to be a limiting factor. Overall, we may expect pTron switching to be on a $10\sim100$ ps timescale based on NbN and related materials' parameters.

Since the pTron memory device is non-volatile, there is no dissipation unless data is read, written or erased. The intrinsically low switching energy ($<$ 2.2 fJ)[6] of the CCM devices is comparable with the SC device drivers. Readout is non-dissipative in the LO state if the R current is correctly chosen, e.g., if $430\ \mu\mathrm{A} < J_{\mathrm{R}} < 480\ \mu\mathrm{A}$ for the device in Fig. 4e. A possibility of reducing the power which has not yet been explored is by reducing the threshold voltage of the CCM, for example by reducing the contact resistance of the Au-$TaS_2$ junctions. In the Hi state, the contact resistance $R_{\mathrm{c}}$ (typically $\sim 100\ \Omega$)[36] is negligible, but the $R_{\mathrm{LO}}$ state resistance can be significantly reduced if $R_{\mathrm{c}}$ can be made smaller through different choice of contact metal or the use of a buffer layer. This will have the effect of reducing the shunt resistance, pushing the pTron more in the direction of the PSC regime. The measured $V - J$ curves agree well with the simulated results, apart from the absence of the hysteresis. This happens because the bias current in the experiment is pulsed as opposed to a DC current used in the simulations. Pulsed bias was intentionally chosen to prevent heating of the CCM device. We observe that external noise can strongly affect the operation of the pTron device, especially when coupled to the extremely sensitive choke terminal. This can cause the pTron to be triggered at much lower bias currents than desirable, leading to too small amplification for proper operation and switching. Proper electromagnetic shielding and filtering is thus very important.

**Conclusions**

We demonstrated that SC NW nTrons and CCM devices are a good match both in terms of impedance and switching thresholds to allow full control of memory operations by tuning of device parameters. The key parameters for achieving a sufficiently large voltage across the NW itself are the critical current $J_{\mathrm{c}}$ and normal state resistance $R_{\mathrm{N}}$. To achieve a sufficiently high voltage $J_{\mathrm{c}}R_{\mathrm{N}} \sim 0.5$ V across the SC NW, the critical current density $j_{\mathrm{c}}$ should be between $\sim 5 - 10\mathrm{MA/cm^2}$, and the normal state sheet resistance of the film should be $R_{\mathrm{N}} \sim 100 - 500\ \Omega/\mathrm{sq}$, depending on the dimensions of the channel of the nTron. Such parameters are easily achievable with present NbN or NbTiN thin film technology[23,25,37,38], and were met with the films used in this work.

Our work paves the way for high-speed operation of CCM and hybrid pTron devices, and towards the possibility of writing data with single SFQ pulses with energy of $E_{\mathrm{W}} \simeq 10^{-20}$ J. Combining our approach with the use of low-loss SC microstrip lines that allow ballistic transfer of the picosecond SFQ signals[9] could lead to the development of near-THz data memory write operations. Implementation that follows established cross-bar array device architectures[39] may be used for making large number of densely packed devices on a chip. Further reduction of energy dissipation may be achieved by optimization of shortening of the QP thermalization time in the SC device and lowering switching thresholds of CCM devices. The size scalability of CCMs and nTrons allows for very flexible design of memory circuits with few-nanometre feature size components. The immediate challenges for advancing this technology lie in the co-fabrication of devices that combine 1*T*-$TaS_2$ CCM and SC NW devices on the same chip.

## Methods

nTron drivers were fabricated out of a magnetron sputtered SC NbTiN thin film (thickness 15 nm) with sheet resistance $R_{\mathrm{sheet}}$ ~108 Ω/sq, critical temperature $T_{\mathrm{C}}\sim 10$ K and critical current density $J_{\mathrm{crit-dens}}\sim$ 6.8 MA/cm$^2$. First, metallic pads for electrical contacting were fabricated using electron beam lithography and liftoff procedure (10 nm Ti/100 nm Au). The rest of the circuit was then fabricated by spin coating hydrogen silsesquioxane (HSQ) resist, patterning with e-beam writer, and etching the NbTiN film with reactive ion etching procedure. To fabricate CCM devices we use electron beam lithography and lift-off to pattern metallic electrodes (5 nm Ti/100 nm Au) over the 1*T*-$TaS_2$ crystal flakes (thickness between 30-80 nm) deposited on Si/SiO2 substrates. nTron and CCM devices were electrically connected with aluminium wire bonds. All the experiments were performed in an Oxford Spectromag cryostat, at different temperatures from 1.5 K up to 7 K.

**Acknowledgments**. We thank for financial support from the Slovenian Research Agency (grants P1-0040, J7-3146, I0-0053, QUASAR), Slovene Ministry of Education, Science and Sport (C3330-19-952005), Cooperation of Science and Technology COST CA16218, and ERC PoC (GA767176). J.R. has received funding from the European Union's Horizon 2020 research and innovation program under the Marie Skłodowska-Curie grant PSI-FELLOW II-3i (GA 701647). NbTiN SC film was prepared at the Quantum Device Lab of the Department of Physics of ETH Zurich. This project has received funding from the European Union's Horizon 2020 research and innovation programme under grant agreement No 101007417 having benefited from the access provided by Paul Scherrer Institute in Villigen, Switzerland within the framework of the NFFA-Europe Pilot Transnational Access Activity, proposal ID286.

**Present Addresses.**
[#]Laboratory for Nano and Quantum Technologies (LNQ), Paul Scherrer Institute, CH-5232 Villigen, Switzerland

## Author contributions

A.M., V.V.K. and D.M. performed the numerical simulations. A.M., R.V., V.Y., B.B., T.L., M.M. performed electrical transport measurements. D.S. and A.M. fabricated CCM devices. J.R., D.K., S.G. and Y.E. fabricated nTron devices. M.G. grew the NbTiN thin film. D.M., I.V. and A.M. devised the experiment.

## Competing interests

The authors declare no competing interests.

FIGURES

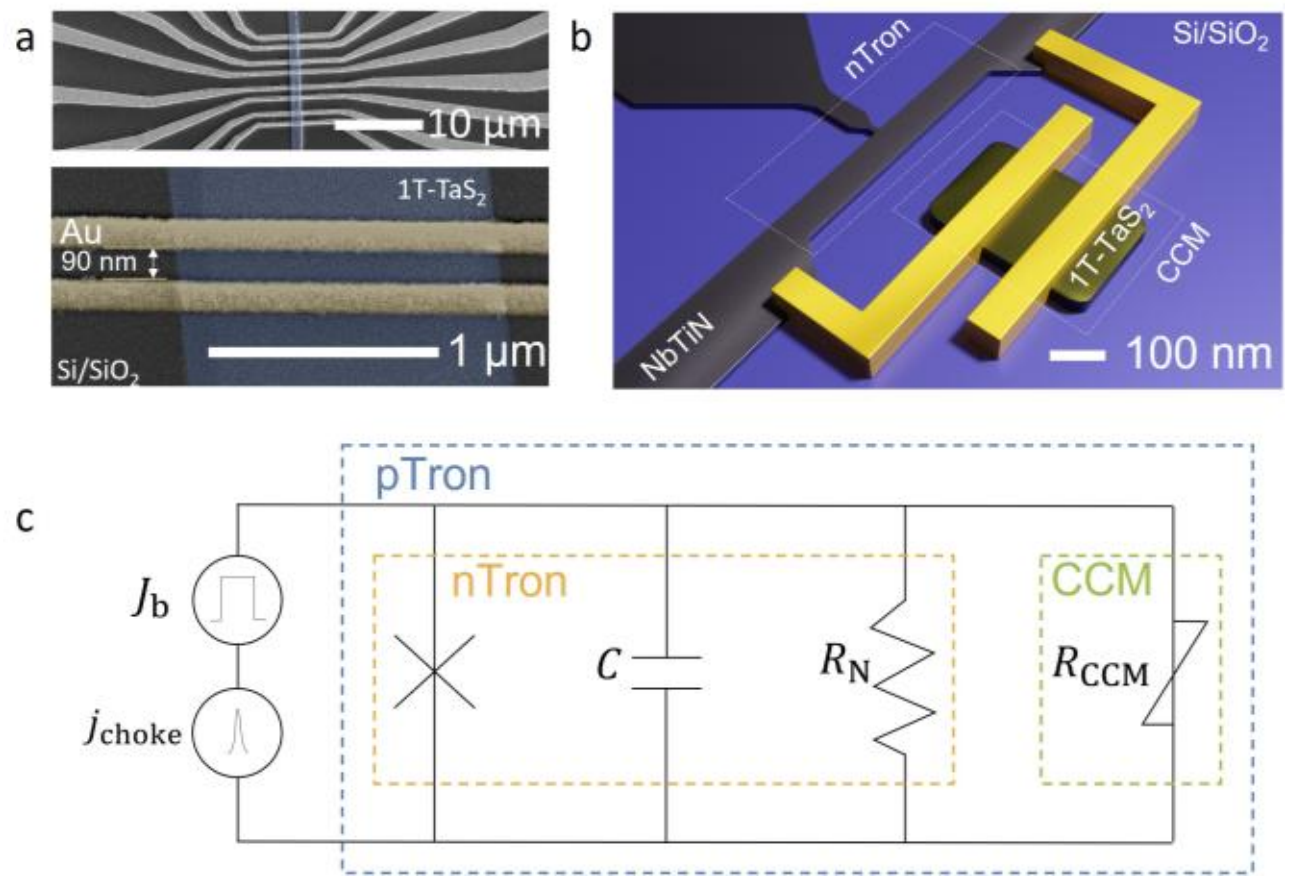

**Figure 1. The pTron device.** a) Scanning electron microscope images of the CCM device with four bits in the top panel, and a zoom-in on one of the bits in the bottom panel. The thickness of the $1T$-$TaS_2$ layer is 80 nm. Its fabrication process is described in Ref.[4]. b) pTron rendering showing individual nTron and CCM devices. c) An equivalent circuit for describing the pTron includes a Josephson junction with a normal state resistance $R_N$ and capacitance $C$ (orange dashed box). The pTron (blue dashed box) includes a switchable CCM shunt resistance $R_{CCM}$ (green dashed box) in parallel with the NW or nTron. The model circuit includes a bias current source $J_b$ and a pulsed source $j_{choke}$ for writing data.

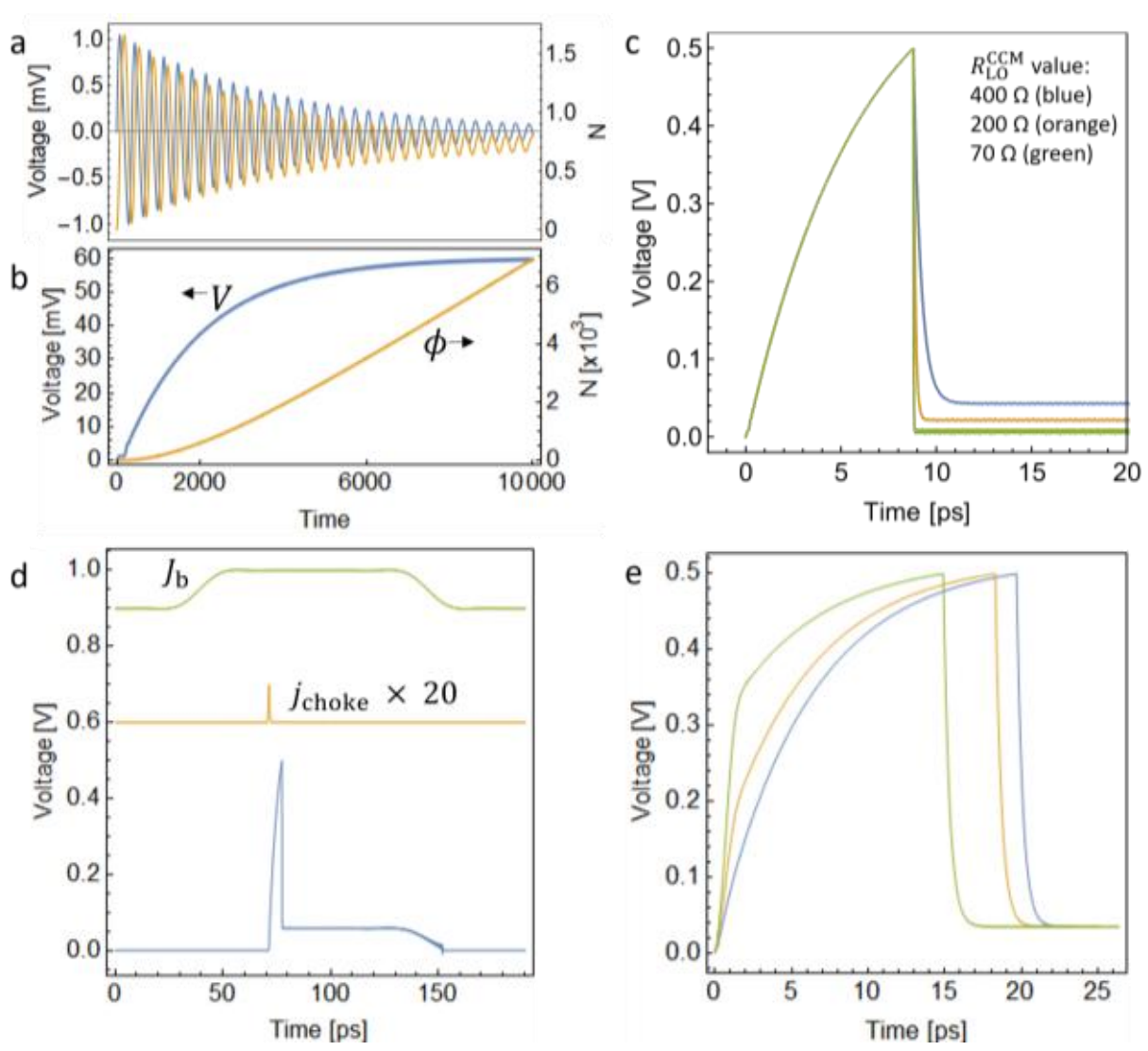

**Figure 2. Simulation of temporal pTron switching dynamics with DC and 1 ps SFQ-pulses.** a) Simulated dynamical response of the phase $\phi$ (orange) and voltage $V = J_b R_T \, d\phi/dt'$ (blue) following an instantaneous increase in current $J_b$ at $t' = 0$: a) below SC NW switching threshold ($J_b = 100$ µA $< J_{sw}$) and b) above threshold ($J_b = 130$ µA $> J_{sw}$). In both cases $R_N = 600\ \Omega$, $R_{CCM} = 2$ kΩ, $J_c = 150$ µA. c) Switching with $J_b = 87$ µA DC current. Here $R_N = 20$ kΩ, $R_{HI}^{CCM} = 8.5$ kΩ, for different $R_{LO}^{CCM} = 400\ \Omega$ (blue), 200 Ω (orange) and 70 Ω (green). d) Voltage across pTron $V$ (blue) after switching by a 1 ps, $j_{choke} = 5$ µA choke pulse (orange) synchronized with a $\tau_B = 100$ ps, $J_b = 87\ \mu A$ bias pulse (green). The rising and falling edges of the bias pulse were rounded with a Fourier filter. e) Switching by different $j_{choke} = 10$ µA (blue), 100 µA (orange) and 250 µA (green) pulses, respectively. Here again $R_N = 20$ kΩ, $R_{HI}^{CCM} = 8.5$ kΩ, $R_{LO}^{CCM} = 400\ \Omega$, and the choke pulse length $\tau_{choke} = 1$ ps.

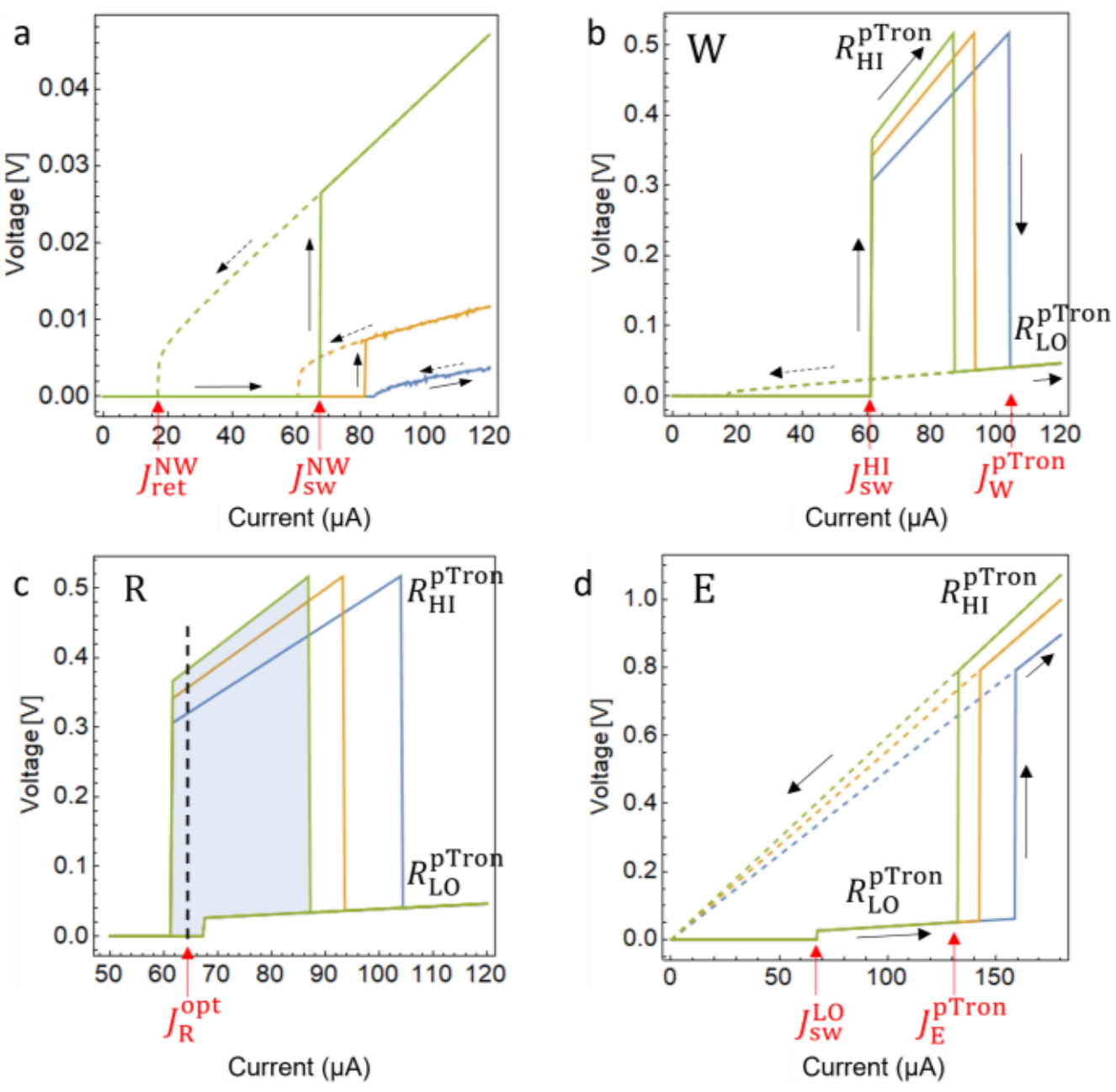

**Figure 3. $V - J$ curves for pTron operations** calculated using Eq. (1) corresponding to the NW, and for W, E and R cycles of the pTron. a) The $V - J$ curves of the NW for different shunt resistances 400 (green), 100 (orange) and 40 Ω (blue), respectively. The non-hysteretic PSC regime is for 40 Ω. b) The W cycle using $J_W^{CCM} = 61\ \mu A$ shown for three different values of $R_N = 12$ kΩ (blue), 16 kΩ (orange) and 20 kΩ (green). Other common circuit parameters are: $R_{HI}^{CCM} = 8.5$ kΩ, $R_{LO}^{CCM} = 400\ \Omega$, $C = 1$ fF, and $J_c = 120$ µA. The return cycle is shown with dashed lines. c) The R cycle. The shaded area indicates where the HI and LO states can be discerned. Note that $R_{LO} = 0$ between 60 and 68 µA, so the ratio $R_{HI}/R_{LO}$ is infinite. Optimal R current is marked. d) The E cycle with $J_E^{CCM} = 93$ µA and for the same three $R_N$ as in b). The return cycle is shown with dashed lines.

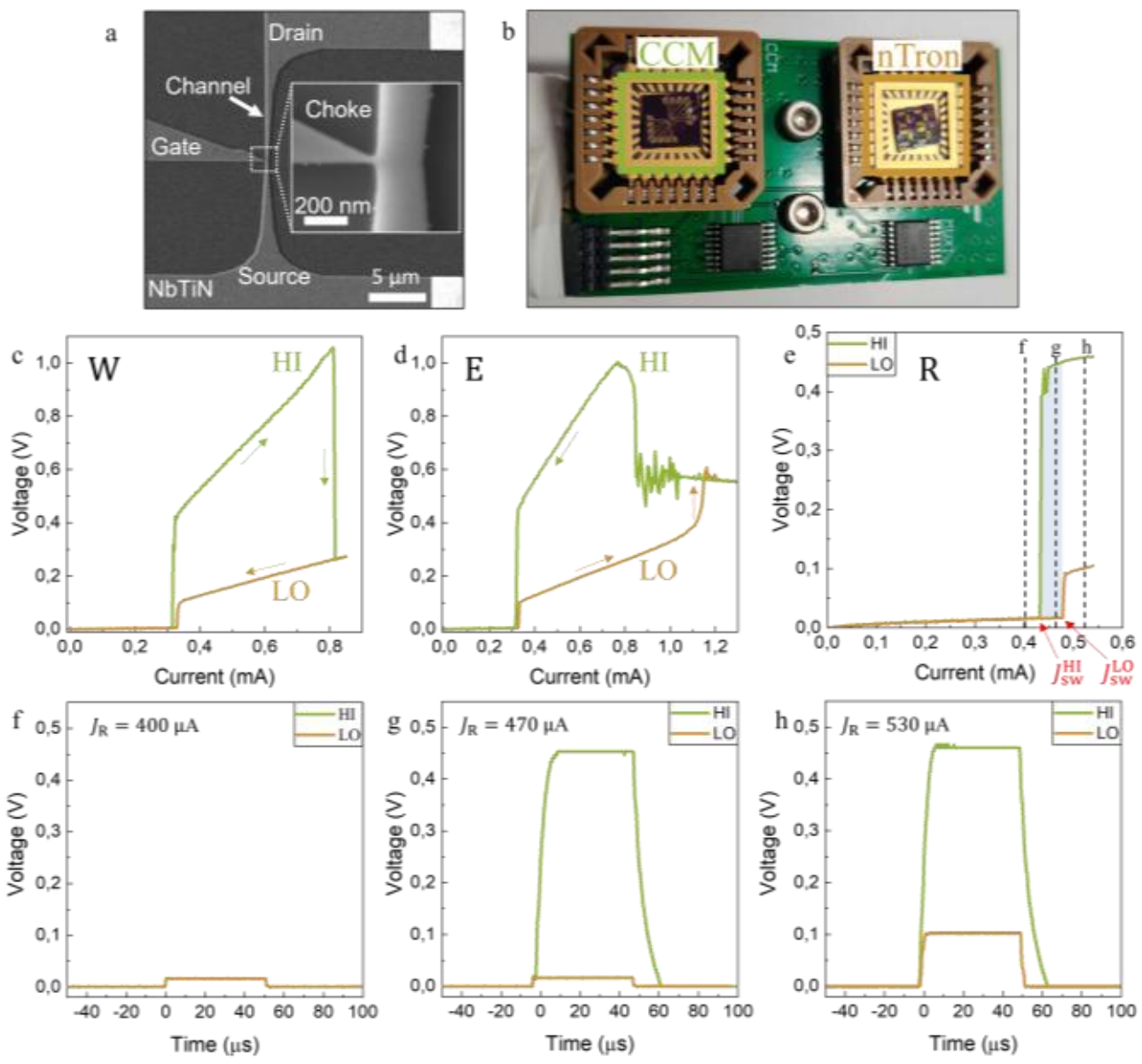

**Figure 4**. **Experimental results on pTron devices.** a) SEM image of a fabricated nTron out of a SC NbTiN thin film. The inset shows the gate terminal that is connected to the channel via a choke constriction. b) Experimental setup with CCM and nTron chips on a low-temperature printed circuit board. c) Measured $V-J$ curve for pTron W operation, where $R_{\mathrm{N}}$=2.1 kΩ, $R_{\mathrm{HI}}^{\mathrm{CCM}}$ =21.7 kΩ, $R_{\mathrm{LO}}^{\mathrm{CCM}}$ = 460 Ω. d) Measured $V-J$ curve for pTron E operation, where $R_{\mathrm{N}}$ is the same, $R_{\mathrm{LO}}^{\mathrm{CCM}}$ =460 Ω and $R_{\mathrm{HI}}^{\mathrm{CCM}}$ =26.7 kΩ. e) Read operation $V-J$ curves for HI and LO state of the pTron. Three different R currents used for real-time measurements shown in panels f, g, h are marked with dashed lines. Shaded blue area shows the optimal R currents. f) Voltage vs. time waveform for $J_{\mathrm{R}}$ =400 µA and $\tau_{\mathrm{R}}$ =50 µs, where both HI and LO states read the same value, zero volts. Note that a non-zero value appears due to a two-point contact. g) R operation with $J_{\mathrm{R}}$ =470 µA. For LO state we read the same as before, while the HI state reads a non-zero value. h) R operation with $J_{\mathrm{R}}$ =530 µA, where both states read non-zero value.